\documentclass[12pt]{article}
\usepackage{amsfonts,amssymb}
\usepackage{graphicx}
\newcommand{\beq}{\begin{eqnarray}}% can be used as {equation} or {eqnarray}
\newcommand{\eeq}{\end{eqnarray}}
\newcommand{\half}{\frac12}

\newcommand{\wg}{\wedge}

\begin{document}
\begin{flushright}%\vspace{-2cm}
{\small
UPR-1120-T, NSF-KITP-05-33 \\ %\vspace{-0.35cm}
hep-th/0506021}%\\
\end{flushright}

\centerline{\Large \bf Mesons and Flavor on the Conifold} \vskip 0.5 cm
\renewcommand{\thefootnote}{\fnsymbol{footnote}}
\centerline{{\bf Thomas S. Levi$^a$} \footnote{tslevi@sas.upenn.edu} {\bf and} {\bf Peter Ouyang$^b$}
\footnote{pouyang@vulcan.physics.ucsb.edu} } \vskip .5cm \centerline{\it $^a$Kavli Institute for Theoretical Physics, University of
California} \centerline{\it Santa Barbara, CA 93106-4030} \centerline{\it and} \centerline{\it David Rittenhouse Laboratories, University of
Pennsylvania} \centerline{\it Philadelphia, PA 19104} \vskip 0.5cm \centerline{\it $^b$Department of Physics, University of California}
\centerline{\it Santa Barbara, CA 93106-9530}

\setcounter{footnote}{0}
\renewcommand{\thefootnote}{\arabic{footnote}}

\begin{abstract}
We explore the addition of fundamental matter to the Klebanov-Witten field theory. We add probe D7-branes to the ${\cal N}=1$ theory
obtained from placing D3-branes at the tip of the conifold and compute the meson spectrum for the scalar mesons. In the UV limit of massless
quarks we find the exact dimensions of the associated operators, which exhibit a simple scaling in the
large-charge limit. For the case of massive quarks we compute the spectrum of scalar mesons numerically.
\end{abstract}

\section{Introduction}

The gauge theory/string theory correspondence \cite{jthroat,US,EW} furnishes a powerful set of tools for understanding gauge
theories at strong coupling by performing computations in a dual string theory at weak coupling.  However, the correspondence is only
well-understood in systems where the string background is highly symmetric and nearly flat, but we expect that the duals to many
interesting gauge theories (such as large-$N$ QCD or SQCD) will not have these properties.  It is therefore an interesting challenge to
study less symmetric string backgrounds, and in particular to study backgrounds with reduced supersymmetry.

One interesting class of models arises from compactifications of string theory on noncompact Calabi-Yau manifolds with D3-branes at conical
singularities, which generically give rise to $\mathcal{N}=1$ gauge theories with product gauge groups and bifundamental matter. These
models are attractive for several reasons.  They possess minimal supersymmetry and are therefore closer to realistic gauge theories than the
well studied $\mathcal{N}=4$ case; also, they lead to conformal field theories where the quantum conformal invariance is not obvious by
inspection of the field theory (but where the supergravity dual makes conformal invariance manifest.)  Perhaps the most striking feature of
these theories is that one can break conformal invariance in a controlled way by adding fluxes through cycles of the Calabi-Yau geometry,
which induce RG flow and confinement at low energies.

However, one missing element of these models is fundamental matter. Aside from being experimentally important, fundamentals give rise to
many interesting things such as the phase structure of super-Yang-Mills theory in the infrared.  In confining theories the fundamentals of
course do not appear as asymptotic states but are instead confined in mesons and baryons.

In this note we study the mesonic fluctuations of a particular set of mesons in the conifold theory of Klebanov and Witten \cite{KW}.  This theory is
interesting for its relative simplicity and also because its non-conformal version flows to a theory very similar to $\mathcal{N}=1$ pure
glue theory in the infrared.  Moreover all metrics for the corresponding supergravity solutions are known, allowing explicit computations.
The mesons which we study arise as fluctuations on D7-branes which are embedded in the string background. The fundamental fields come from
strings connecting the stack of D3-branes to the D7-branes.  In the usual decoupling limit, the 3-7 strings and 3-3 strings, which describe
the gauge theory, have a dual description in terms of the closed strings and 7-7 strings.  The closed strings are the usual glueballs of the
strongly coupled field theory while the open 7-7 strings are naturally identified with the mesons.

We will compute the spectrum of operator dimensions, which, as we will see, can be done exactly for a large portion of the states, and we will
study the effect of giving masses to the quarks (which requires numerical work).

The paper is organized as follows. In section \ref{sec-rev} we review the geometry of the conifold. In section \ref{sec-flavor} we discuss
adding flavor to the Klebanov-Witten field theory by the addition of probe D7-branes. In section \ref{sec-scalar} we compute the spectrum
for scalar mesons. In the case of massive quarks, we compute the mass spectrum numerically, but in the massless case (corresponding to the UV limit of the gauge theory) we obtain the spectrum analytically.  In section \ref{sec-con} we discuss our results.

\section{Review of the Conifold} \label{sec-rev}

In this section we briefly review the geometry of the conifold in order to fix notation.  Useful references are
\cite{KW,coco,MP,MT,Ceres,Ohta}.

The conifold is a non-compact Calabi-Yau 3-fold, defined by the equation
\beq z_1 z_2 -z_3 z_4 =0 \label{coneq} \eeq
in ${\bf C}^4$. Because Eqn.(\ref{coneq}) is invariant under an overall real rescaling of the coordinates, this space is a cone, whose base is the
Einstein space $T^{1,1}$ \cite{KW,coco}. The metric on the conifold may be cast in the form \cite{coco}
\beq ds_6^2 = dr^2 + r^2 ds_{T^{1,1}}^2\ , \label{conimetric} \eeq
where
\begin{equation} \label{co}
ds_{T^{1,1}}^2= {1\over 9} \bigg(d\psi + \sum_{i=1}^2 \cos \theta_i d\phi_i\bigg)^2+ {1\over 6} \sum_{i=1}^2 \left( d\theta_i^2 + {\rm
sin}^2\theta_i d\phi_i^2
 \right)
\
\end{equation}
is the metric on $T^{1,1}$. Here $\psi$ is an angular coordinate which ranges from $0$ to $4\pi$, while $(\theta_1,\phi_1)$ and
$(\theta_2,\phi_2)$ parametrize two ${\bf S}^2$s in the standard way. This form of the metric shows that $T^{1,1}$ is a $U(1)$ bundle over
${\bf S}^2 \times {\bf S}^2$.

These angular coordinates are related to the $z_i$ variables by
\beq
z_1 &=& r^{3/2} e^{i/2(\psi-\phi_1-\phi_2)}\sin\frac{\theta_1}{2}\sin\frac{\theta_2}{2}, \nonumber \\
z_2 &=& r^{3/2} e^{i/2(\psi+\phi_1+\phi_2)}\cos\frac{\theta_1}{2}\cos\frac{\theta_2}{2},\label{ztoangles}\\
z_3 &=& r^{3/2} e^{i/2(\psi+\phi_1-\phi_2)}\cos\frac{\theta_1}{2}\sin\frac{\theta_2}{2}, \nonumber\\
z_4 &=& r^{3/2} e^{i/2(\psi-\phi_1+\phi_2)}\sin\frac{\theta_1}{2}\cos\frac{\theta_2}{2}.\nonumber  \eeq
It is also sometimes helpful to consider a set of ``homogeneous'' coordinates $A_a, B_b$ where $a,b=1,2$, in terms of which the $z_i$ are
\beq
z_1&=&A_1 B_1, \qquad z_2=A_2 B_2, \\
z_3&=&A_1 B_2, \qquad z_4=A_2 B_1. \label{ztoAB} \eeq
With this parameterization the $z_i$ obviously solve the defining equation of the conifold.

We may also parameterize the conifold in terms of an alternative set of complex variables $w_i$, given by
\beq
\begin{array}{ll}
z_1 = w_1 + iw_2,& \qquad z_2 = w_1 -iw_2, \\
z_3 = -w_3+iw_4,& \qquad z_4 = w_3 +i w_4.
\end{array}
\label{ztow} \eeq
The conifold equation may now be written as
\beq \sum w_i^2 =0 \label{coneqw} \eeq
and we identify the $T^{1,1}$ base of the cone as the intersection of the conifold with the surface
\beq \sum |w_i|^2 = r^3. \eeq
$T^{1,1}$ described in this way is explicitly invariant under $SO(4)\simeq SU(2)\times SU(2)$ rotations of the $w_i$ coordinates and under
an overall phase rotation.  Thus the symmetry group of $T^{1,1}$ is $SU(2)\times SU(2) \times U(1)$.

An important fact about $T^{1,1}$ is that it has Betti numbers $b_2,b_3=1$.  The corresponding two-cycle and three-cycle may be expressed in
terms of harmonic differential forms:
\beq
\omega_2 &=& \half \left( \Omega_{11} -\Omega_{22}\right) ,\\
\omega_3 &=& \zeta \wg \omega_2 .
\eeq

In this paper we will consider D7-branes in the model of Klebanov and Witten \cite{KW}.  This model is a particularly simple ${\cal N}=1$
gauge/gravity dual, obtained by placing a stack of $N$ D3-branes near a conifold singularity.  The branes source the RR 5-form flux and warp
the geometry:
\beq
ds_{10}^2 &=& h(r)^{-1/2} dx_{\mu} dx^{\mu} + h(r)^{1/2}(dr^2+r^2 ds_{T^{1,1}}^2) \\
h(r) &=& 1+ \frac{L^4}{r^4}\\
g_s F_5 &=& d^4x \wg dh^{-1} + \star ( d^4x \wg dh^{-1})\\
L^4 &=& \frac{27}{4} \pi g_s N \alpha'^2. \eeq
Hereafter, we specialize to the near-horizon limit $r/L \ll 1$, and set $L=1$ for convenience. It may be easily restored by dimensional analysis at any point.

The dual field theory has gauge group $SU(N) \times SU(N)$ and matter fields $A_{1,2},B_{1,2}$ which transform in the bifundamental color
representations $({\bf N},{\bf \bar{N}})_c$ and $({\bf \bar{N}},{\bf N})_c$.  The theory also has a superpotential
\beq W=\lambda Tr(A_i B_j A_k B_l) \epsilon^{ik} \epsilon^{jl}. \eeq
By solving the F-term equations for this superpotential, we obtain supersymmetric vacua for arbitrary diagonal $A_{1,2}$ and $B_{1,2}$, so
that the moduli space of the field theory is precisely that of $N$ D3-branes placed on the conifold.

\section{Adding flavor} \label{sec-flavor}

In this section we review the procedure of adding flavor branes to AdS/CFT in general and make several useful comments on adding flavor to
the Klebanov-Witten field theory both in terms of the bulk geometry and the dual field theory. This general procedure was first pointed out
in \cite{AFM,Karch,karchrandall} and was exploited in the $AdS_5 \times S^5$ case in \cite{myers}.  Some other examples of flavored theories with probe branes have been studied in \cite{bertd7,granad7,Naculich,Nastase,Sakai,Sakai2,Barbon,Nunez,Erdmenger,Evans,Erlich,Babington,Wang}.

One way to add flavor to AdS/CFT is to take a system of D3-branes and then to add D7-branes which fill the four $x^\mu$ directions and four
of the six transverse dimensions \cite{Karch}.  In flat space such a configuration of branes is clearly supersymmetric.  As usual there is
an ${\cal N}=4$ $SU(N)$ SYM theory living on the D3-branes. Strings with one end on a D3-brane and one end on a D7-brane couple to the
fields of the D3-brane gauge theory as quarks.

For AdS/CFT purposes we can now take the supergravity approximation in which D3-branes are replaced by an $AdS_5 \times S^5$ geometry with
Ramond-Ramond flux, while we retain the D7-branes as probes which fill the five AdS directions and which wrap a topologically
trivial 3-cycle of the internal 5-manifold (for example an $S^3$ submanifold of the $S^5$ of $AdS_5 \times S^5$).
% (in our case this will be the space $T^{1,1}$).
The triviality of the 3-cycle guarantees that the brane carries no net charge and will not introduce any tadpoles.  On the other hand,
topological triviality also suggests that one might be able to shrink the $S^3$ and slip it off of the $S^5$, naively in contradiction with
the flat space picture of D3 and D7-branes.
 It turns out that subtleties of the $AdS$ geometry play a key role in ensuring stability.  The mass eigenvalues of modes controlling the
D-brane slipping off the 3-cycle are negative, but are above the Breitenlohner-Freedman bound \cite{BF}, so that the 7-brane
embedding is stable.

In the flat space picture, if the D3-branes and D7-branes intersect then the quarks are massless, and if the D3-branes and D7-branes are
separated then the quarks are massive.  This translates nicely into the AdS picture in the following way.  A D7-brane which intersects the
D3-branes in flat space gets mapped to a D7-brane which fills the whole AdS space and wraps a three-sphere of constant size in the $S^5$.
On the other hand, a D7-brane separated from the stack of D3-branes maps to a D7-brane which wraps an $S^3$ with some asymptotic size at
large AdS radius, but this $S^3$ shrinks to zero size at some finite radius (which is possible because of the topological triviality).  In the
5-dimensional AdS space the D7-brane appears to fill out the radial direction up to some minimal radius where it ``ends.''

It is interesting of course to consider theories with branes in spaces which are not flat.  The basic picture of D3 and D7 branes
contributing gauge fields and quarks will not change, but many details are different.  For simplicity throughout this paper we specialize to
the case of a single D7-brane.  If the number of D3-branes is large then the D7 backreaction can be systematically neglected and it is
appropriate to treat the D7-brane as a probe, which we do throughout this paper. Inclusion of backreaction effects in other geometries has
been explored in \cite{AFM,pz,Kirsch}.

Let us consider D7-branes embedded in the geometry of the conifold by the equation $z_1=\mu$.  In terms of the standard coordinate system,
\beq
z_1=r^{3/2} e^{i/2(\psi-\phi_1-\phi_2)}\sin\frac{\theta_1}{2}\sin\frac{\theta_2}{2}
\nonumber
\eeq
so the embedding equation gives two conditions, one on
the magnitude of $z_1$ and one on the phase:
\beq
r_0 &=& \left(\frac{|\mu|}{\sin\frac{\theta_1}{2}\sin\frac{\theta_2}{2}}\right)^{2/3} , \label{rembed} \\
\psi_0 &=& \phi_1+\phi_2 +{\rm const}. \eeq
This embedding can be explicitly shown to be supersymmetric by considering
the $\kappa$-symmetry on the worldvolume of the brane \cite{conprobes}.  A slightly different embedding equation was studied in the warped deformed conifold by \cite{Kuperstein}.

%In the field theory we have added fundamental matter and cancellation of gauge anomalies requires that we have two flavors in each gauge
%group with opposite chiralities. 

It was proposed in \cite{peter} that the embedding $z_1=\mu$ leads to fields, summarized in Table \ref{quarktable} and a superpotential of the form
\beq
W &=& W_{flavors}+W_{masses}, \\
W_{flavors} &=& h \tilde{q} A_1 Q+ g q B_1 \tilde{Q}, \ \  W_{masses} = \mu_1 q \tilde{q} + \mu_2 Q \tilde{Q}. \eeq
To relate this superpotential to the D7-brane geometry, let us probe the space
with a single D3-brane, which corresponds to giving some expectation values to $A_1$ and $B_1$.  One then finds that the theory on this
probe has a massless flavor when $A_1B_1 = \mu_1\mu_2/(gh)$, which is exactly of the form of the embedding equation $z_1=\mu$.  Part of the motivation for this superpotential was a comparison with a type IIA brane construction \cite{Park} where a D6-brane splits on an NS5-brane, contributing two flavor branes and correspondingly two sets of flavors.  For the type IIB picture, in the massless limit of the field theory, this corresponds nicely to the presence of two solution branches of $z_1=0$, namely $\theta_1=0$ and $\theta_2=0$.  If the quarks are massless there is an $SU(K)\times SU(K)$ flavor symmetry, where $K$ is the number of
probe D7-branes. If the quarks are massive then the two branches of the D7-branes connect and the flavor symmetry is broken down to the diagonal $SU(K)$. 

\begin{table}
\begin{center}
\begin{tabular}{||c|c|c||} \hline
Field& $SU(N_c)\times SU(N_c)$& $SU(K)\times SU(K)$ \\ \hline \hline $q$& $({\bf N},{\bf 1})$& $({\bf K},{\bf 1})$\\ \hline $\tilde{q}$&
$({\bf \bar{N}},{\bf 1})$&$({\bf 1},{\bf K})$\\ \hline $Q$& $({\bf 1},{\bf N})$& $({\bf 1},{\bf \bar{K}})$\\ \hline $\tilde{Q}$& $({\bf
1},{\bf \bar{N}})$& $({\bf \bar{K}},{\bf 1})$\\ \hline
\end{tabular}
\end{center}
\caption{Representation structure of the added $\mathcal{N}=1$ flavors.}\label{quarktable}
\end{table}

An alternative perspective is to suppose that one of the masses $\mu_i$ is larger than the other and then integrate out the associated
flavors.  Then one obtains a quartic superpotential of the form
\beq W = q(A_1 B_1 - \mu) \tilde{q} \eeq
which again produces the appropriate massless locus for a D3-brane probe.  Our probe calculations will show that this quartic superpotential is consistent with adding D7-branes with massive flavors (and then with the limit where we take the masses to zero.)  Of course, because we believe the quarks can be massive the consistency was virtually guaranteed.  However, taking the limit $\mu\rightarrow 0$ and setting $\mu = 0$ are different things, and it is unclear whether the theory corresponding to the cubic superpotential can be realized or not.

\section{Scalar mesons} \label{sec-scalar}
In this section, we compute the dimension and mass spectra of the scalar mesons. As discussed in the introduction, in the probe and
decoupling limits the 7-7 strings are identified with the mesons in the dual field theory. We will thus be able to extract the mass spectrum
of the spin=0 mesons and their conformal dimension in the UV limit by studying the 7-7 strings.

The semiclassical dynamics of this D7-brane are captured by the Dirac-Born-Infeld action
\beq S_{DBI} = \tau_7 \int d^8\xi \sqrt{-\det_{ij}\left(g_{MN} + F_{MN}\right)\frac{\partial y^M}{\partial \xi^i}\frac{\partial
y^N}{\partial \xi^j}} +\frac{g_s \tau_8}{2} \int C_4 \wg F_2 \wg F_2. \eeq
where $\xi^i$ are coordinates on the D7-brane.  We will compute the spectrum of fluctuations for the D7-branes using this action.

Let us consider the fluctuations of scalar modes alone, with all D7-brane gauge fields turned off.  Then the DBI action is simply the
worldvolume of the 7-brane.  Let us choose as coordinates on the brane eight of the spacetime coordinates: $(x^{\mu},
\theta_1,\theta_2,\phi_1,\phi_2).$  The fluctuations can be described by setting
\beq
r&=&r_0(\theta_i)(1+\chi(x^{\mu},\theta_i,\phi_j)) ,\\
\psi&=&\psi_0(\phi_i) + 3\eta(x^{\mu},\theta_i,\phi_j). \eeq
The unperturbed induced metric takes the form
\beq g_{MN}&=& \left(
\begin{array}{ccc}
r_0^2\eta_{\mu\nu}& 0 &0  \\
0 & g_{\theta_i\theta_j} & 0\\
0 & 0 & g_{\phi_i \phi_j}
\end{array} \right) \label{wholemetric} \\
g_{\theta_i \theta_j}&=& \left(
\begin{array}{cc}
\frac16 +\frac19 \cot^2\frac{\theta_1}{2}& \frac19 \cot\frac{\theta_1}{2}\cot\frac{\theta_2}{2}\\
\frac19 \cot\frac{\theta_1}{2}\cot\frac{\theta_2}{2}&\frac16 +\frac19 \cot^2\frac{\theta_2}{2}
\end{array} \right) \label{thetaparts} \\
g_{\phi_i \phi_j} &=& \left(
\begin{array}{cc}
\frac16\sin^2\theta_1+\frac19(1+\cos\theta_1)^2&\frac19(1+\cos\theta_1)(1+\cos\theta_2)\\
\frac19(1+\cos\theta_1)(1+\cos\theta_2)&\frac16\sin^2\theta_2+\frac19(1+\cos\theta_2)^2
\end{array}
\right) \label{phiparts} \eeq
One expands about this metric via the matrix identity
\beq \sqrt{\det{A+\delta A}} = \sqrt{\det A} (1+ \frac12 {\rm Tr} A^{-1}\delta A + \frac18({\rm Tr} A^{-1}\delta A)^2-\frac14 {\rm Tr}
A^{-1}\delta A A^{-1}\delta A + ... \eeq
The terms first order in the fluctuations $\chi$ and $\eta$ turn out to be total derivatives, as is necessary for our embedding to be a
solution of the equations of motion.  The quadratic order fluctuations lead to an action of the form
\beq \label{scalaraction}
S &=& \tau_7 \int d^4x d\theta_1 d\theta_2 d\phi_1 d\phi_2 \sqrt{\det g_0}\frac{1}{C}\Bigg[\frac12 g_0^{ab}\partial_a \chi \partial_b \chi+\frac12 g_0^{ab}\partial_a \eta \partial_b \eta \nonumber \\
&+& \frac{4}{\sin^2\frac{\theta_2}{2}} \chi \partial_{\phi_2} \eta  -\frac{2}{C\sin^2\frac{\theta_2}{2}}\left(\cot\frac{\theta_1}{2}\partial_{\theta_1}+\cot\frac{\theta_2}{2}\partial_{\theta_2}\right)\chi\partial_{\phi_2}\eta\\
&+& \frac{4}{\sin^2\frac{\theta_1}{2}} \chi \partial_{\phi_1}
\eta-\frac{2}{C\sin^2\frac{\theta_1}{2}}\left(\cot\frac{\theta_1}{2}\partial_{\theta_1}+\cot\frac{\theta_2}{2}\partial_{\theta_2}\right)\chi\partial_{\phi_1}\eta
\Bigg] \nonumber \eeq
with
\beq C=1+\frac23 \cot^2\frac{\theta_1}{2}+\frac23 \cot^2\frac{\theta_2}{2}. \eeq

\subsection{The UV/massless limit}

Even though the quarks have mass, if we flow to the UV, they effectively become massless and conformal symmetry is restored, at least at the
classical level. It is interesting to inquire what the dimensions of the operators are in the UV field theory. In the dual, this corresponds
to computing near the boundary of $AdS$.

Examining (\ref{rembed}) we see that there are two ways to go near the boundary: $\theta_1 \to 0$ or $\theta_2 \to 0$. Which one we choose will
determine which side of the conifold we are on near the boundary. In the current
setup, the physics is symmetric between exchange of $\theta_1$ and $\theta_2$ so we will simply choose the $\theta_1 \to 0$ limit. 
We can compute the dimensions of the operators in the field theory by examining the scaling of the 7-7 strings near the boundary.

We define $r_0=\mu^{2/3} e^{-\beta/3}$ (now a good coordinate because of the conformal invariance), and $\cos^2\frac{\theta_2}{2} = x$.  Defining the linear combination of fields
\beq \label{lincombo} \Phi^{\pm} = \chi \pm i \eta \eeq
we find that the equations for $\Phi^\pm$ are two fully decoupled partial differential equations.  This equation is solved by a separation of
variables ansatz
\beq \Phi ^\pm= \rho^{\pm}(x) e^{k\beta/3}e^{im_1 \phi_1 + im_2\phi_2}. \eeq
The equations of motion for the scalar reduces to ordinary differential equations for the functions $\rho(x)^\pm$ which take the form
\beq \Bigg((1-x)\frac{\partial}{\partial x} x \frac{\partial}{\partial x} + \frac16 k(k-1)-\frac{m_1^2}{8}\frac{3-x}{1-x} + \frac{m_1
m_2}{2(1-x)}& &\nonumber \\ -\frac14 \frac{m_2^2}{x(1-x)}\mp \frac{m_1}{4} \frac{1+x}{1-x}\pm \frac{m_2}{2(1-x)}\Bigg)\rho^{\pm}(x) &=& 0.
\label{eqnrho} \eeq
This equation has singularities only at $x=0,1,\infty$ and is therefore of hypergeometric type.  To see this explicitly we define new
functions by rescaling the $\rho^{\pm}$ by factors of $x$ and $(1-x)$, which allow us to remove the terms in (\ref{eqnrho}) proportional to
$\frac{1}{x}$ and $\frac{1}{1-x}$.  Explicitly, we write
\beq \rho^{\pm}(x) &=& x^{p} (1-x)^{q} f^{\pm}(x) \eeq
for which the equation of motion becomes
\beq \Bigg[x(1-x)\frac{\partial^2}{\partial x^2} + \left(1+2p -x(1+2p+2q)\right)\frac{\partial}{\partial x}+\frac{(q^2-q + \frac14
-\frac14(m_1-m_2\pm 1)^2)}{1-x}\nonumber \\ \hspace{-1cm}+\frac{(p^2-\frac{m_2^2}{4})}{x} +\frac16 k(k-1)- \frac14 \left(\frac{m_1^2}{2}\mp
m_1 \right) -\left(p+q\right)^2\Bigg]f^{\pm}(x) = 0. \eeq
Appropriate choice of the parameters $p$ and $q$ eliminates the terms proportional to $1/x$ and $1/(1-x)$.  The wavefunctions are given by
\beq f^{\pm}(x) = _2 \! \! F_1(-\alpha, \alpha +2(p+q),1+2p;x) \eeq
which are regular when $\alpha$ is a non-negative integer; it turns out that the original $\rho$ are also regular with this condition over
the range $0\leq x \leq 1$ ($0 \leq \theta_2 \leq \pi$) which encompasses our domain.  We also find that there are two possible values of
$k$:
\beq
k_1 &=& \frac12 + \frac12 \sqrt{1+3m_1^2 \mp 6m_1 +24(\alpha+p+q)^2}\\
k_2 &=& \frac12 - \frac12 \sqrt{1+3m_1^2 \mp 6m_1 +24(\alpha+p+q)^2}. \eeq

To be painfully explicit, we exhibit the solutions for $f^+$ (the $f^-$ are straightforwardly related.)  It is clear that there are always
two choices of $p$ and $q$ which do the trick; to make regularity transparent we will always choose $p$ and $q$ to be positive.  We then
have four cases:
\begin{itemize}
\item $m_2\ge0, m_1\ge m_2$: We choose $p=m_2/2$ and $q= 1 +\frac12(m_1-m_2)$, so that $f^+= _2 \! \!  F_1(-\alpha,2+m_1+\alpha,1+m_2;x)$.  The wavefunction is regular if $\alpha$ is a non-negative integer (negative $\alpha$ gives irregular or redundant solutions) and so the two values of $k$ are quantized to be
\beq k=\frac12\pm \frac12 \sqrt{1+3m_1^2 -6m_1 +24(\alpha+\frac{m_1}{2}+1)^2}. \eeq

\item $m_2\ge0, m_1<m_2$: Again we choose $p=m_2/2$, but now to make regularity obvious we take $q=(m_2-m_1)/2$, such that $q>0$.  Now $f^+= _2 \! \! F_1(-\alpha,2m_2-m_1+\alpha,1+m_2;x)$, again with $\alpha$ a non-negative integer.  The quantized values of $k$ are
\beq k=\frac12\pm \frac12 \sqrt{1+3m_1^2-6m_1+24(\alpha+m_2-\frac{m_1}{2})^2}. \eeq

\item $m_2<0,m_1\ge m_2$: Now we choose $p=-m_2/2$ and $q= 1 +\frac12(m_1-m_2)$, finding that $f^+= _2 \! \! F_1(-\alpha,2+m_1-2m_2+\alpha,1-m_2;x)$, with $\alpha$ a non-negative integer.  The allowed values of $k$ are
\beq k=\frac12\pm \frac12 \sqrt{1+3m_1^2-6m_1+24(\alpha+\frac{m_1}{2}-m_2+1)^2}. \eeq

\item $m_2<0,m_1< m_2$: Now we choose $p=-m_2/2$ and $q=(m_2-m_1)/2$, finding that $f^+= _2\!\! F_1(-\alpha,\alpha-m_1,1-m_2;x)$, with $\alpha$ a non-negative integer.  The allowed values of $k$ are
\beq k=\frac12\pm \frac12 \sqrt{1+3m_1^2-6m_1+24(\alpha-\frac{m_1}{2})^2}. \eeq
\end{itemize}

%Demanding regularity of the solution both at $x=1,0$ ($\theta_1 =0, \pi$) leads to two possible pairs of values for $k$: \beq
%k_1 ^{\pm}&=& \frac12 + \frac12 \sqrt{1+3m_1(m_1\mp2)+24(a+1\pm \frac{m_1}{2})^2} \\
%k_2 ^{\pm}&=& \frac12 -\frac12 \sqrt{1+3m_1(m_1\mp2)+24(a+1\pm \frac{m_1}{2})^2} \eeq
%
%and
%
%\beq
%\tilde{k}_1 ^{\pm} &=& \frac12 + \frac12 \sqrt{1+3m_1(m_1\mp2)+24(a\pm m_2 \mp \frac{m_1}{2})^2} \\
%\tilde{k}_2 ^{\pm} &=& \frac12 -\frac12 \sqrt{1+3m_1(m_1\mp2)+24(a\pm m_2 \mp \frac{m_1}{2})^2} . \eeq 

To find the dimensions of the operators, we recall that  $e^{k\beta/3} \sim r^{-k}$.  In the AdS/CFT correspondence a minimal massless scalar field dual to an operator of dimension $\Delta$ scales as $r^{-\Delta}$ for its normalizable part and $r^{\Delta-4}$ for its non-normalizable part. However, by examining (\ref{scalaraction}) we see that the kinetic terms for these scalars are not canonically normalized, which means that the possible scalings at infinity are modified to  $r^{-\Delta+p}$ and $r^{\Delta-4+p}$ for some $p$. Using the values for $k_1$ and $k_2$ we have $2\Delta-4=k_1-k_2$, which one can compute straightforwardly.  

The dimensions are mostly complicated irrational numbers (reminiscent of the closed string spectrum on $T^{1,1}$ \cite{Ceres,Gubser}) but a few features of the spectrum stand out.  The lowest mode has $m_1=m_2=\alpha=0$, and is simply a constant; it can be assigned dimension 5/2 or 3/2.  From the earlier discussion of the massive field theory, it is natural to choose dimension 3/2 and associate this mode with the operator $q\tilde{q}$.  Note also that the mode with $m_1=m_2=1$ and $\alpha=0$ has dimension 3, appropriate for a superpotential term -- we identify this mode with the operator $qA_2B_2\tilde{q}$.  If added to the superpotential, this operator would change the D7-brane embedding from $z_1=\mu$ to $z_1 +\epsilon z_2 = \mu$.

For large $m_1$, all the dimensions scale as $\Delta \sim
3m_1/2$.  This is consistent with identification of the corresponding gauge theory operators as
\beq q(AB)(AB)\ldots(AB)\tilde{q} \eeq
where, ignoring the $q$ fields, each insertion of $(AB)$ should increase the dimension by 3/2 and the relevant $SU(2)$ charge (associated
with $m_1$) by one unit.  Unlike the case of baryonic operators on the conifold, where one finds an exact scaling $\Delta \sim 3N/4$
\cite{GK}, we see that the mesons only exhibit a simple scaling with the charge in a large-charge limit. For small charges there are
boundary effects due to the quarks which, at least in the large-$N$ limit, are completely calculable here.  This behavior should also be
contrasted with the case of flavors added to $\mathcal{N}=4$ super-YM theory, where the meson dimensions were pure integers.

If we take instead the limit of large $\alpha$, we see that the dimensions scale as $\Delta \sim \sqrt{6}\alpha$.  It would be interesting to find an explanation for this curious scaling in the field theory.

\subsection{The mass spectra}
Having obtained the conformal dimensions for the mesons in the conformal limit, we would like to compute their spectrum by solving the full
differential equation without taking any simplifying limits. Unfortunately, we will find that the equation is not amenable to analytic
solution and so we will have to appeal to numerical methods. We will display selected results from several cases that are illustrative of
the general behavior.

We will again find that the linear combination of fields (\ref{lincombo}) decouples the equations of motion. 
%into two partial differential
%equations. 
Rewriting the action (\ref{scalaraction}) in terms of $\Phi^\pm$ and varying gives
\beq \label{massdiffeq} \frac{1}{\sqrt{-g_0}} \partial_a \biggl(\frac{1}{C} \sqrt{-g_0} g_0^{ab} \partial_b \Phi^\pm \biggr) \pm 3i
\biggl[\frac{1}{\sqrt{-g_0}}
\partial_{\theta_i} (\sqrt{-g_0} \gamma^{\theta_i \phi_j} ) - \gamma^{\phi_j} \biggr] \partial_{\phi_j} \Phi^\pm =0 ,
\eeq
where the $a,b$ indices run over the $x^\mu$ {\it and} the $\theta_{1,2} , \phi_{1,2}$, and
\beq \gamma^{\theta_i \phi_j} &=& \frac{4 (1+\cos \theta_j)}{C^2 \sin^2 \theta_j} \partial_{\theta_i} \ln r_0 (\theta_i,\theta_j) , \\
\gamma^{\phi_j} &=& \frac{8(1+\cos \theta_j)}{3 C \sin^2 \theta_j} . \eeq
The inverse components of the metric are straightforward to find from the form given in (\ref{wholemetric}-\ref{phiparts}). Since
$\partial_{x^\mu}$ and $\partial_{\phi_i}$ are Killing vectors we can write
\beq \Phi^\pm= \psi^\pm (\theta_1,\theta_2) e^{ik \cdot x} e^{im_1 \phi_1 +im_2 \phi_2} . \eeq
We find that (\ref{massdiffeq}) becomes
\beq &-& \partial_{\theta_i} \biggl(\frac{1}{C} \sqrt{-g_0} g_0 ^{\theta_i \theta_j} \partial_{\theta_j} \psi^\pm \biggr) +\frac1C
\sqrt{-g_0} g_0 ^{\phi_i \phi_j} m_i m_j \psi^\pm \pm 3 \biggl[ \partial_{\theta_i} ( \sqrt{-g_0} \gamma^{\theta_i \phi_j} ) - \sqrt{-g_0}
\gamma^{\phi_j} \biggr] m_j \psi^\pm \nonumber \\
 &=& -\frac{\sqrt{-g_0}}{C r_0 ^2 (\theta_1,\theta_2)} k^2 \psi^\pm . \label{massdiffeq2} \eeq
Note that for massive modes $k^2 =k_\mu k^\mu < 0$ since it must be timelike. Using simple Kaluza-Klein arguments we see that the mass of
the mesons in the dual field theory is $M^2=-k^2$. It is evident from the form of this equation that the only difference between the
equation for $\psi^+$ and for $\psi^-$ is in a term proportional to $m_j$. Thus, we can choose to solve for $\psi^+$ without loss of
generality. This equation cannot be solved analytically. In addition, we have found no simple way to separate the equation in the
$\theta_1$,$\theta_2$ directions and so we must use a numerical approach to solving the partial differential equation.

\subsubsection{The numerical approach}

\begin{figure}
\begin{center}
\includegraphics[width=0.5\textwidth]{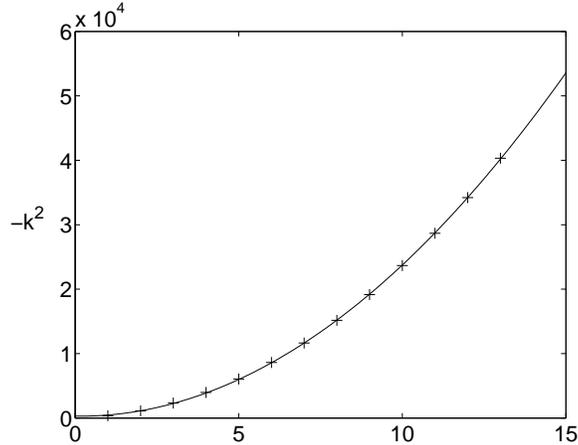}
\end{center}
\caption{Mass eigenvalues versus eigenvalue number for the zero node modes with $m_1=m_2=0$, $\mu^{-4/3}=.02$. $+$s denote actual
eigenvalues and the solid line is a best fit line with equation $-k^2=240n^2-68n+290$.} \label{nolnonodesfig}
\end{figure}

Because we are unable to separate the partial differential equations into ordinary differential equations, we must use a technique slightly
more involved than the regular finite difference scheme and shooting technique. We will make use of the finite element method and the
Arnoldi algorithm via Matlab to solve for the mass eigenvalues, $-k^2$. We will use a mesh with 2779 nodes and 5392 triangles for all
problems involved. The $\theta_i$s are part of two different $S^2$s and thus range from $0$ to $\pi$. Since we already know $\theta_i \to
0$ corresponds to going near the boundary, and we want normalizable modes, we place Dirichlet boundary conditions at $\theta_{1,2}=0$. We
must also demand regularity at $\theta_{1,2}=\pi$, which corresponds to placing Neumann boundary conditions at $\theta_{1,2}=\pi$.

\begin{figure}
\begin{center}
\includegraphics[width=0.5\textwidth]{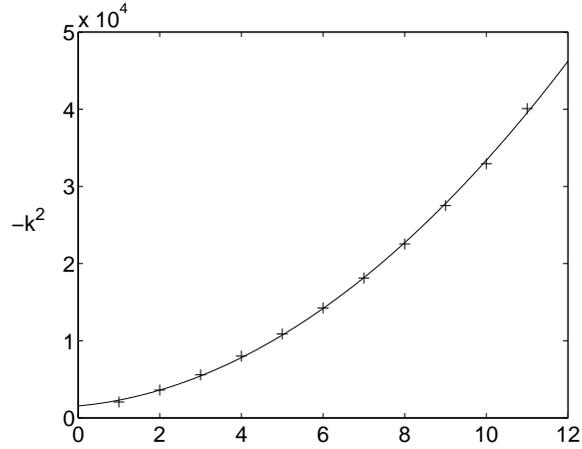}
\end{center}
\caption{Mass eigenvalues versus eigenvalue number for the one node modes with $m_1=m_2=0$, $\mu^{-4/3}=.02$. $+$s denote actual eigenvalues
and the solid line is a best fit line with equation $-k^2=270n^2+490n+1500$.} \label{nolonenodefig}
\end{figure}

We will first examine the simplest case, when $m_1=m_2=0$. Setting $\mu^{-4/3}=.02$ we solve (\ref{massdiffeq2}) for the first 50
eigenvalues. The eigenvalues break up into different series corresponding to the number of nodes in the $(\theta_{1},\theta_{2})$ plane. In
figures \ref{nolnonodesfig} and \ref{nolonenodefig} we display the first two such series. Higher series have similar behavior. The $+$ signs
denote actual mass eigenvalues, while the solid lines are best fit lines.

\begin{figure}
\begin{center}
\includegraphics[width=0.5\textwidth]{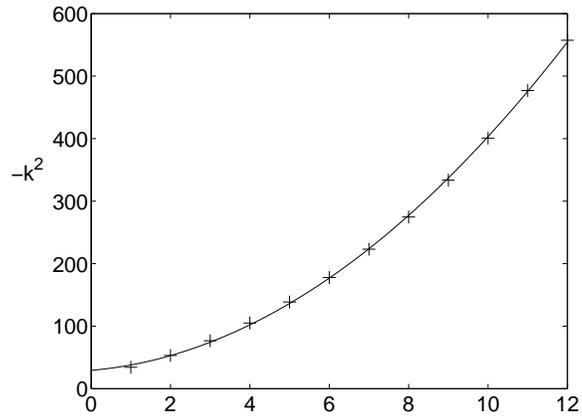}
\end{center}
\caption{Mass eigenvalues versus eigenvalue number for the zero node modes with $m_1=1$, $m_2=2$, $\mu^{-4/3}=2$. $+$s denote actual
eigenvalues and the solid line is a best fit line with equation $-k^2=3.2n^2+5.4n+29$.} \label{l12nonodesfig}
\end{figure}

We will also find similar behavior for modes with $m_{1,2} \neq 0$. In figure \ref{l12nonodesfig} we display the zero node modes for the
case $m_1=1$, $m_2=2$ with $\mu^{-4/3}=2$ (note: changing $\mu$ just changes the eigenvalues by an overall scaling, as expected since it
merely scales the mass gap for the quarks). We find similar behavior for other values of $m_{1,2}$ and different number of nodes.

For all the cases we see that in the large $n$ limit we have $M \sim n \mu^{2/3} $ as we would expect. Restoring $L$ by dimensional
analysis, and using $m_q \sim \frac{\mu^{2/3}}{2 \pi \alpha'}$ we find the the mass gap for the lightest meson is
\beq M_{gap} \sim \frac{m_q}{\sqrt{g_s N}} \eeq
Therefore, in the supergravity regime where $g_s N \gg 1$ we find the meson mass is much smaller than the quark mass. At large t'Hooft
coupling we find that the binding energy of the mesons almost completely cancels the rest energy of the quarks. This is similar to the
situation in $AdS_5 \times S^5$ \cite{myers}.

\section{Discussion} \label{sec-con}

In this note we have computed the spectrum of mesons in an $\mathcal{N}=1$ field theory corresponding to fluctuations in the position of a
holomorphically embedded D7-brane.  In the limit of nearly massless quarks, the field theory is classically conformal, and also conformal at
large-$N$, and the spectrum turns out to be computable exactly, where the dimensions in general are complicated irrationals.

There are a few operators for which the exact results are simple.  Among these are the lowest mode, corresponding to a mass term, with dimension 3/2, and a mode corresponding to a BPS fluctuation of the D7-brane, with dimension 3.  The existence of these operators suggests that a consistent superpotential for our flavored theory is 
\beq W= qA_1B_1\tilde{q}. \eeq
It would be interesting to study the Klebanov-Strassler theory \cite{KS} obtained at the end of the duality cascade with the addition of 3-form flux with this superpotential.

In the strictly massless limit, $z_1=0$, it is possible to relax our embedding condition slightly.  Specifically, with a nonzero mass we imposed a relation between the azimuthal coordinates, $\psi -\phi_1-\phi_2=0$.  However, when the mass is zero this condition need not apply; it would be nice to see what relaxing this condition would mean for the field theory (in particular, whether it is possible to realize the cubic superpotential discussed in section 3.)

The appearance of irrational dimensions is not surprising, in light of similar results for the glueball spectrum of the conifold
\cite{Ceres,Gubser}.  However, this feature of the meson spectrum differs from the $\mathcal{N}=4$ case, where the meson dimensions were
pure integers.  In particular, we do not find a tower of states with spacing 3/2, except in the large R-charge limit; more precisely, in
this limit the spectrum is of the form $3k/2 + O(1/J)$.  It might be possible to compute these $1/J$ corrections in a plane-wave limit, or perhaps in some other formalism.  It would be interesting if such a comparison with our exact results were possible.

We have also numerically computed the spectrum for the case of massive quarks. In the large $g_s N$ limit the meson mass
gap is significantly smaller than the quark masses. We have uncovered a relatively simple quadratic scaling behavior for the meson masses.
It would be nice to find, either with analytical or more numerical work, the exact functional dependence on $n, m_1, m_2$ etc. 

All of our calculations have been in the probe limit and further studies of the backreaction would be interesting, especially for the
Klebanov-Strassler deformed conifold theory.  However, it may still be possible to learn things from further study of probe theories.  In
particular, it would be interesting to study the dynamics of nontrivial classical field configurations in the D7-brane worldvolume.  Such
fields would correspond to dissolved D3-branes or anti-D3-branes.  The anti-brane case is particularly interesting, as it would break
supersymmetry along the lines of the KKLT scenario\cite{KKLT}\footnote{We thank S.~Trivedi for this suggestion.}, but with the possibility for some
moduli to be fixed by the D7-brane.  We leave these suggestions for the future.

\vspace{0.2in} {\centerline {\bf Acknowledgments}}

We thank A.\ Maharana for discussions and collaboration during the early stages of our work. We thank H.\ Elvang, E.\ Katz, M.\
Strassler, S.\ Trivedi, and J.~Wacker for discussions and correspondence. We are especially grateful to Joel Giedt and Leo Pando-Zayas for bringing some typos in a previous version to our attention.  TSL thanks the Kavli Institute for Theoretical Physics for warm hospitality during much of the
preparation of this work. TSL was supported in part by the KITP under National Science Foundation grant PHY99-07949, the National Science
Foundation under grants PHY-0331728 and OISE-0443607, and the Department of Energy under grant DE-FG02-95ER40893.   The work of P.~O. is supported in part by the DOE under grant DOE91-ER-40618 and by the NSF under grant PHY00-98395.

%\bibliographystyle{utphys}
%\bibliography{mesonref}
\providecommand{\href}[2]{#2}\begingroup\raggedright\endgroup

\end{document}